**Cross-individual generalizability of machine learning models for ball speed prediction in baseball pitching**

Ryota Takamido[1], Chiharu Suzuki[1], Hiroki Nakamoto[2]

1 Sports Innovation Organization, National Institute of Fitness and Sports in Kanoya, Kanoya, Kagoshima 891-2393, Japan

2 Faculty of Physical Education, National Institute of Fitness and Sports in Kanoya, Kanoya, Kagoshima 891-2393, Japan

Corresponding Author: Ryota Takamido

Street Address, Kanoya, Kagoshima, 891-2393, Japan

Email address: rtakamido@nifs-k.ac.jp

**Abstract**

Although machine learning (ML)-based performance outcome prediction is an important topic in contemporary sports science, one important issue is the limited understanding of the cross-individual generalizability of ML models in sports contexts. To address this issue, this study aimed to evaluate the cross-individual generalizability of ML models for predicting ball speed in baseball pitching. A dataset comprising 50 pitchers from various competitive levels was analyzed. Cross-individual generalizability was assessed using leave-one-subject-out cross-validation. Specifically, the effects of expertise level and restrictions on spatiotemporal motion information were examined to identify factors influencing model generalizability. The results revealed that, under cross-individual evaluation, (1) predictive performance was markedly lower than under within-individual evaluation, with $R^2$ decreasing from 0.91 to 0.38; (2) the model tended to overestimate the performance of Intermediate pitchers relative to Expert pitchers, with a significant group difference in signed prediction error ($p < .05$); and (3) the trunk and pivot leg demonstrated relatively high generalization performance, with the pivot leg showing notable generalizability even during the weight-shift initiation phase ($R^2 > 0.25$). These findings underscore the importance of cross-individual evaluation in enhancing the practical applicability of ML in sports settings and contribute to a deeper understanding of the biomechanical factors underlying the target movement.

## Introduction

Sports movements are characterized by complex inter-joint coordination and long kinematic chains extending from the lower limbs through the trunk to the upper limbs. Consequently, the relationship between movement patterns (i.e., how an athlete moves) and performance outcomes is often nonlinear and cannot be adequately captured by a simple linear model (Middleton et al., 2015).

Machine learning (ML) models offer a powerful approach for modeling such complex and nonlinear relationships. Unlike conventional statistical methods that typically assume linear relationships between movement variables and performance outcomes (e.g., greater knee flexion leading to increased jump height), ML models can capture more complex patterns in the data. For instance, convolutional neural networks (CNNs) can capture spatial coordination among multiple joints (Li et al., 2017), while recurrent neural networks (RNNs) can model long-term temporal dependencies within kinematic chains (Liu et al., 2018). Owing to these advantages, ML models have been increasingly used to predict performance outcomes from motion data (Jia et al., 2025), demonstrating high predictive accuracy across various sports tasks, including ball type and speed prediction in baseball pitching (Takamido et al., 2026; Yang et al., 2025), tennis shot direction prediction (Shimizu et al., 2019), and penalty kick direction prediction (Freire-Obregón et al., 2026).

Despite these advances, the cross-individual generalizability of ML models in sports remains insufficiently understood. Athletic movements involve complex coordination across multiple joints and exhibit large inter-individual variability (Giles et al., 2023; Müller et al., 2015). Therefore, it is essential to assess whether ML models can accurately predict outcomes for unseen individuals using validation approaches such as leave-one-subject-out cross-validation (LOSOCV). Such evaluation enhances the potential for real-world application and helps identify movement characteristics that are consistently associated with performance across individuals.

However, many previous studies on ML applications in sports have either included data from the same athletes in both the training and test sets or failed to clearly describe data partitioning across individuals, as highlighted by Worsey et al. (2021). Similar limitations have been observed in recent studies on performance prediction (Chang et al., 2026; Manzi et al., 2024). As noted in prior work (Gholamiangonabadi et al., 2020; Worsey et al., 2021), evaluating ML models only at the individual level may lead to overestimation of predictive performance. Although some studies (e.g., Link et al., 2022; van den Tillaar et al., 2021) have employed LOSOCV to validate predictive or classification performance, several important questions remain underexplored.

(1) *Does skill level affect the cross-individual generalizability?* If experts exhibit greater inter-individual variability than novices and intermediates (Takamido et al., 2019), cross-individual generalization may be more difficult for ML models. Moreover, the greater movement efficiency of experts (Burns et al., 2020) may influence the direction of prediction errors. That is, if skilled athletes achieve higher performance than others, despite similar motion inputs, ML models may consistently underestimate expert performance.

(2) *Which body segments and movement phases affect cross-individual generalizability?* If the movements of specific body segments at specific phases commonly contribute to performance across individuals, the presence or absence of such information should substantially affect the predictive accuracy of ML models.

A detailed examination of these points would not only deepen our understanding of the target movement but also offer the potential to generate new insights and hypotheses in a data-driven manner.

Accordingly, this study aimed to examine the cross-individual generalizability of ML models for performance outcome prediction from the above two perspectives. Specifically, ball speed prediction for baseball pitching was selected as the target task for the ML model. This selection was made for the following reasons. First, an effective long-term kinematic chain among whole-body joints is required to increase ball speed (Howenstein et al., 2019). This increases inter-individual variability and makes cross-individual generalization more difficult. Moreover, the effective kinematic chain of skilled pitchers contributes to a greater movement efficiency (Crotin et al., 2022), which may influence the direction of prediction errors. Although Yang et al. (2025)

demonstrated the potential of ML models for accurate prediction in a ball speed prediction task, the procedure for separating data across players was not described in detail, and cross-individual generalizability was not examined in detail.

## 2. Materials and Methods

### 2.1 Dataset development

The dataset comprised baseball pitching motions and corresponding ball velocity data collected from 50 pitchers and was identical to that used in our previous study (Takamido et al., 2025). The participants included four high school players, 10 collegiate athletes, 20 industrial league players, three independent league players, and 13 professional league players.

Each athlete performed more than five fastballs under simulated game conditions. Pitching motions were recorded using 16 optical motion capture cameras (Raptor-E; Motion Analysis Corporation, Santa Rosa, CA, USA) at a sampling rate of 200–500 Hz. The ball velocity, defined as the initial speed, of each pitch was also recorded as the performance index using the TrackMan Baseball system (TrackMan, Vedbæk, Denmark). For analysis, the top five fastballs were selected for each athlete based on ball velocity. The positions of 15 anatomical landmarks were extracted from the selected data, including the parietalis (head), bilateral acromions (shoulder), lateral epicondyle of the humerus (elbow), radial styloid process (wrist), greater trochanter of the femur (hip), lateral condyle of the femur (knee), and heel and top of the shoes (toe). Data from left-handed pitchers were mirrored to match the orientation of right-handed pitchers.

Raw motion data were preprocessed using a fourth-order Butterworth filter, with the optimal cutoff frequency determined according to the method proposed by Schreven et al. (2015). Ball release time was defined as the moment at which the wrist velocity of the throwing arm reached its maximum. A fixed-duration segment was then extracted from each motion sequence, spanning from 1.0 s before to 0.2 s after ball release, covering the weight-shift, release, and follow-through phases. The three-dimensional positions of the 15 joints were temporally normalized to 101 frames, resulting in an input matrix of $15 \times 101 \times 3$ (joints × time frames × feature variables) for each pitch.

The data were previously collected for educational and training purposes. Accordingly, an opt-out consent framework was adopted in accordance with ethical guidelines. Information about the study, including its purpose and use, was made publicly available to ensure transparency. Individuals whose data were included had a clear opportunity to decline participation. The study procedures were approved by the Institutional Ethics Committee of the National Institute of Fitness and Sports, Kanoya (approval number: 25-1-26). All the procedures adhered to the principles of the Declaration of Helsinki.

The source code and anonymized dataset used in this study are publicly available at GitHub (https://github.com/takamido/ Ball_speed_pred_cross_ind).

## 2.2 Cross-individual generalizability analysis

### 2.2.1 Baseline model selection

Using the developed dataset, we evaluated the cross-individual generalizability of ML models for predicting ball speed in baseball pitching. Figure 1 provides an overview of the analytical framework. As shown in Figure 1a, the baseline ML model was first selected for use throughout the study. This approach ensured that the effects of motion input and model architecture or parameter settings could be examined independently, without confounding effects arising from simultaneous changes.

Specifically, the baseline model was defined as one that achieved the highest predictive accuracy for cross-individual predictions using whole-body information across all time points. The Graph Neural Network–Recurrent Unit (GNN-GRU) model (Yang et al., 2025) and Transformer (Vaswani et al., 2017) were selected as candidate model architectures. The GNN-GRU model showed the best predictive performance in a previous study on ball speed prediction (Yang et al., 2025). Although that study did not clearly describe how the dataset was partitioned, we included this model as a candidate to evaluate the cross-individual generalization performance. The Transformer is a representative model for time-series analysis that uses self-attention mechanisms to model sequential data, enabling it to capture long-term dependencies within time-series data, including position-tracking data (AlShami et al., 2023) and motion data (Zhao et al., 2026) in the sports context. Because a previous study (Yang et al., 2025) did not include the transformer in its comparison with the GNN-GRU model, it was also selected in this study as another candidate model, given its strong capacity for modeling time-series data.

To account for the effects of the hyperparameter settings, six models with different parameter configurations were prepared for each architecture (Table 1). For GNN-GRU, the number of GNN layers and hidden units varied based on the parameter settings used in a previous study (Yang et al., 2025). The Transformer consisted of three stacked transformer encoders, and the number of attention heads, latent dimension ($d_l$), and feedforward network dimension ($d_f$) were varied as hyperparameters. Because this study focuses on the application of these models, detailed descriptions of their architectures are omitted; readers can refer to previous studies (Vaswani et al., 2017; Yang et al., 2025).

All the models were trained and evaluated using the following LOSOCV procedure: Specifically, all the samples from one pitcher were used as the test set, whereas the samples from the remaining pitchers were used for training. This procedure was repeated for all the pitchers, with each pitcher serving as the test subject. During the training process, the model was trained on each training set using the Adam optimizer with a learning rate of 0.001 and a weight decay of $1\times10^{-4}$. Mean squared error was used as the loss function. The training was performed for up to 50 epochs. An early stopping criterion was introduced based on the coefficient of determination ($R^2$) of the training set; training was terminated when $R^2$ exceeded 0.90. This criterion was adopted to prevent overfitting, while allowing the model to adapt sufficiently to the training data. Consequently, all training runs were terminated early according to this criterion. The training was conducted on a Google Colab platform using an NVIDIA T4 GPU.

After each training session, the trained model was used to predict the ball velocity of the five pitches thrown by the pitcher assigned to the test set, and its predictive accuracy was evaluated. Based on the test results, the mean predicted value for each of the 50 pitchers was calculated for each model. These values were used to compute the coefficient of determination ($R^2$) with respect to the actual mean value for each pitcher, and the best-performing model was selected.

Finally, for the selected baseline model, an additional dataset was constructed for comparison with the within-individual validation, in which one of the five pitches for each of the 50 pitchers was randomly assigned to the test set. This resulted in a training set of 200 pitches and test set of 50 pitches from the same pitchers. The selected baseline model was trained and evaluated on this dataset using the same procedure.

### 2.2.2 Analysis 1: Effects of expertise on cross-individual generalizability

Using the selected baseline model, the effect of expertise on cross-individual generalizability was evaluated using the selected ML model (Figure 1b). First, each of the 50 pitchers was assigned to either the Expert or Intermediate group. Based on the mean ball velocity data of pitchers across the five competition levels included in this study, we determined the grouping that maximized the following equation:

$$\eta = \frac{d_{ie}}{\sigma_i + \sigma_e}, \tag{1}$$

where $d_{ie}$ denotes the difference between the mean ball velocities of the two groups, $\sigma_i$ and $\sigma_e$ represent the within-group standard deviations of the Intermediate and Expert groups, respectively. This equation was used to determine the grouping that maximized the difference between the two groups in terms of ball velocity while considering within-group variability.

After testing all 15 possible combinations of competition levels (i.e., ${}_5C_1+{}_5C_2$), high school and collegiate pitchers were classified into the Intermediate group (*n* = 14), whereas corporate league, independent league, and NPB pitchers were classified into the Expert group (*n* = 36). The mean ball velocity (± SD) was 77.60 ± 4.41 for Intermediate group and 84.54 ± 4.39 mph for Expert group.

Subsequently, the effects of expertise on cross-individual generalizability were analyzed. First,

we examined whether expertise affected the magnitude of the ML prediction errors. Specifically, the mean absolute error of the baseline model was calculated for the players in each group, and the group-wise means and standard deviations were computed. An independent-samples *t*-test was conducted to determine whether the absolute error differed significantly between groups. The significance level was set at 5%, and the effect size (Cohen's *d*) was calculated. If expert pitchers exhibit greater inter-individual variability and are, therefore, more difficult to generalize across individuals, they would show larger absolute errors.

Furthermore, to examine whether the direction of ML prediction errors exhibited group-specific tendencies, the same procedure was applied to signed prediction errors. As discussed in the introduction, if the greater movement efficiency of expert pitchers causes an underestimation of ball velocity, prediction errors (i.e., predicted values minus true values) would show a negative bias. These analyses were conducted using MATLAB (R2025b, MathWorks).

### 2.2.3 Analysis 2: Spatial and temporal factors influencing cross-individual generalizability

This analysis aimed to identify spatial and temporal features that enhance cross-individual generalizability. To achieve this, spatiotemporal input data were systematically restricted, and the resulting changes in predictive accuracy were examined (Figure 1c). For spatial analysis, body joints were grouped into five regions: throwing arm (right shoulder, elbow, and wrist in the mirrored data), leading arm (left shoulder, elbow, and wrist), trunk (head, both shoulders, and both hips), pivot leg (right hip, knee, heel, and toe), and leading leg (left hip, knee, heel, and toe). Overlap was permitted at joint connections (i.e., shoulders and hips).

For the temporal aspect, 10 datasets per region were generated by progressively increasing the time window in increments of 10 frames (Figure 1c). This design examined the point at which the generalization performance of the ML model began to improve. In other words, we intended to determine when and in which body regions information related to ball velocity emerged across individuals. Although several previous studies (Bourne et al., 2011; Takamido et al., 2026) employed a sliding-window approach with a fixed window length, this study adopted a cumulative time-window design, in which the number of included time frames progressively increased. This approach was selected to ensure that information from the ongoing kinematic chain was always retained.

In total, 50 combinations (five regions × 10 time spans) of spatiotemporal motion were analyzed, and the corresponding datasets were constructed. The training and evaluation conditions for each dataset were the same as those used for the baseline model selection. To account for randomness in the model training, training and evaluation under each condition were repeated 10 times, and the mean and standard deviation of the resulting performance metrics were calculated for each body region and time span. A total of 25,000 model trainings and tests (five regions × 10 time spans × 50 pitchers × 10 iterations) were conducted, and the results are summarized accordingly.

## 3. Results

### 3.1 Baseline model selection

Table 2 summarizes the predictive performance of each model architecture across parameter settings. The Transformer model with four attention heads, $d_l$ = 64 and $d_f$ = 128 exhibited the highest generalization performance ($R^2 = 0.38$) and was therefore selected as the baseline model for subsequent analyses.

In terms of predictive performance, when the baseline model was evaluated using the within-individual dataset split, it achieved an $R^2$ of 0.91, which was comparable to the value reported in the previous study by Yang et al. (2025). Therefore, a substantial decline in predictive performance was observed when the prediction was performed across individuals rather than within individuals ($R^2$ decreasing from 0.91 to 0.38). Figure 2 presents the prediction results of the baseline model for individual pitchers under cross-individual evaluation.

### 3.2 Analysis 1: Effects of expertise on generalizability

Figure 3 presents the results of Analysis 1. Figure 3(a) presents the mean and standard deviation of the absolute prediction error for the Expert and Intermediate groups, while Figure 3(b) presents the corresponding values for the signed prediction error. The independent-samples t-test revealed no significant difference in absolute error between the groups ($t(48) = 0.12$, $p = .89$, $d = 0.04$).

However, a significant difference was observed in signed prediction error (predicted values minus true values) between the groups ($t(48) = 2.13$, $p = .03$, $d = 0.67$). Therefore, although the Expert group did not show a clear directional bias in prediction errors, the Intermediate group exhibited more positive prediction errors (i.e., overestimation).

### 3.3 Analysis 2: Spatial and temporal factors influencing generalizability

Figure 4 presents the mean values for each body segment–phase combination, along with the corresponding standard deviation ranges. Overall, generalization performance was relatively high for the pivot leg and trunk and lowest for the leading arm. Regarding the temporal aspects, while the trunk, leading leg, and leading arm showed a tendency for generalizability to increase over time toward ball release, the pivot leg and throwing arm showed a U-shaped pattern. These results suggest that cross-individual generalizability differs depending on the body segment and movement phase.

## 4. Discussion

The results of this study provide important insights into ML-based performance outcome prediction. First, the baseline model tests showed a substantial decline in predictive performance under cross-individual evaluation compared with within-individual evaluation, with $R^2$ decreasing from 0.91 to 0.38. This result is consistent with previous studies (Gholamiangonabadi et al., 2020; Worsey et al., 2021). which reported reduced predictive performance in cross-individual generalization owing to the substantial inter-individual variability inherent in human movement. Although a previous study reported a 15% decline in accuracy for a daily activity prediction task (Gholamiangonabadi et al., 2020), the present study, which focused on the pitching motion, showed a more pronounced decline. This finding suggests that cross-individual generalization may be more difficult for athletic movements involving complex, long-term, whole-body kinematic chains. Therefore, cross-individual evaluation is important for assessing the utility of ML applications intended for practical sports settings with previously unseen individuals.

Additionally, this study examined the factors that influence the cross-individual generalizability of the ML models in more detail. The results of Analysis 1 suggested that the ML model tended to overestimate ball speed in intermediate pitchers (Figure 3). There are two possible explanations for this observation. The first is the difference in movement efficiency. As shown in a previous study (Howenstein et al., 2019), expert pitchers exhibited a more effective kinematic chain that enabled them to throw balls faster and more efficiently. Therefore, even when similar motion kinematics are observed, the ML model may overestimate the ball speed in intermediate pitchers whose underlying movement efficiency is lower than that of expert pitchers in the training dataset. This suggests that, as indicated in several previous studies (e.g., Teranishi et al., 2023), it may be possible to assess player-specific ability based on the extent to which an individual's actual performance deviates from the ML prediction.

A second explanation is bias toward the center of the target distribution. ML models often produce fewer predictions at the extremes of the target distribution and tend to favor values closer to the center, particularly under standard regression objectives (Ribeiro & Moniz, 2020). Although the plot in Figure 2 does not appear to show a clear deviation from the line with a slope of 1, such a bias may have influenced the results. Although such a bias may be reasonable for reproducing human perception with ML (Körding & Wolpert, 2004), from the perspective of developing a robust ML model, the potential factors that may increase the prediction error should be addressed for further model improvements. If a larger sample size were available in future studies, a more appropriate approach would be to compare players within the same ball-speed range.

Although the relative contributions of these factors remain unclear, the findings highlight the need to account for expertise-related bias when developing ML models applicable across diverse skill levels.

Analysis 2 further revealed that cross-individual generalizability varies across body regions and time spans. The high cross-individual generalizability observed for the trunk and pivot leg suggests that these regions contribute to ball speed in a manner that is consistent across individuals. Previous studies have identified these regions as key components of the kinematic chain: trunk rotation accelerates the subsequent motions of the elbow and wrist (Aguinaldo & Escamilla, 2019), while a strong push-off of the pivot leg accelerates the body in the forward direction (Kageyama et al., 2014). In contrast to distal regions such as both arms, which may vary largely across individuals, the roles of the pivot leg and trunk may have been shared across pitchers with higher and lower ball speeds.

Another key finding from Analysis 2 is the temporal aspect of cross-individual generalizability. To better understand the high generalizability observed in the early phase of the movement, particularly for the pivot leg, Figure 5 shows example postures of the (a) Intermediates and (b) Experts at the 20% time point, corresponding to the first peak in the pivot leg. As shown in the figure, the ML model may capture subtle differences in posture at the onset of weight transfer. From our observations, some intermediates ( left and center examples in Figure 5(a)) appeared to have initiated forward motion of the upper body at this time point, with the hip located further in front of the knee than in the experts. This early forward movement of the upper body may lead to earlier shoulder rotation, potentially interfering with hip–shoulder separation, which is one of the key factors contributing to high ball velocity (Sgroi et al., 2015). Although many previous studies have focused on the later pitching phase after lead-foot contact, this result suggests that common characteristics of high- or low-speed pitchers may also exist in the earlier phases of motion. However, as shown by the example on the right in Figure 5a, some intermediate pitchers were similar to the experts, and a more detailed investigation is needed in future hypothesis-driven analyses.

Overall, examining generalization performance from both spatial and temporal perspectives not only enables a more detailed evaluation of ML models but also provides a deeper understanding of the nature of the target skill.

This study has several limitations. First, the analysis was limited to ball speed prediction in baseball pitching; future studies should extend cross-individual evaluations to other sports to identify sports-specific difficulties and challenges in ML applications. Second, the sample size ($n$ = 50) may limit the model's ability to capture inter-individual variability; larger datasets could improve generalization performance. Third, the study relied on relatively simple input features (i.e., time series of three-dimensional joint positions), and incorporating more advanced feature representations may improve the generalization performance. Finally, although the contribution of each body region was examined through ablation analysis, a more detailed interpretation of the results is needed through the introduction of explainable AI techniques (Minh et al., 2022).

## 5. Conclusion

In conclusion, the main findings of this study can be summarized as follows. Under cross-individual evaluation, (1) predictive performance decreases substantially under cross-individual evaluation, with $R^2$ decreasing from 0.91 to 0.38; (2) ML models tend to overestimate the performance of Intermediate pitchers relative to Experts; and (3) the trunk and pivot leg exhibit high generalization performance, with the pivot leg exhibiting a certain level of generalizability, even during the weight-shift initiation phase ($R^2 > 0.25$). Despite its limitations, this study highlights the importance of cross-individual evaluation for improving the practical applicability of ML in sports settings and for advancing the understanding of the biomechanical determinants of performance.

## Acknowledgments

### Funding

This work was supported by the Japan Society for the Promotion of Science (Grant number: JP25K21018). The funders had no role in study design, data collection and analysis, decision to publish, or preparation of the manuscript.

## Declaration of interest statement

The authors report there are no competing interests to declare

## Data availability statement

The source code and anonymized dataset used in this study are publicly available at GitHub (https://github.com/takamido/ Ball_speed_pred_cross_ind).

**Tables**

**Table 1.** Parameter settings of the candidate models used in the baseline test.

| Base architecture<br>(Hyper parameters) | Parameter settings for baseline test |
|---|---|
| GNN-GRU<br>(GNN Layers, Hidden units) | (2, 32), (2, 64), (2, 128), (3, 32), (3, 64), (3, 128) |
| Transformer<br>(Attention heads, , ) | (2, 32, 64), (2, 64, 128), (4, 64, 128)<br>(4, 128, 256), (8, 128, 256), (8, 256, 512) |

**Table 2**. Results of the baseline model selection test.

| Base Architecture (Hyper parameters) | Parameter settings | Cross-individual generalizability ($R^2$) |
|---|---|---|
| GNN-GRU (GNN Layers, Hidden units) | (2, 32) | 0.14 |
| | (2, 64) | 0.20 |
| | (2, 128) | 0.28 |
| | (3, 32) | 0.21 |
| | (3, 64) | 0.25 |
| | (3, 128) | 0.28 |
| Transformer (Attention heads, $d_l$, $d_f$) | (8, 256, 512) | 0.17 |
| | (8, 128, 256) | 0.18 |
| | (4, 128, 256) | 0.31 |
| | **(4, 64, 128)** | **0.38** |
| | (2, 64, 128) | 0.21 |
| | (2, 32, 64) | 0.27 |

## Figures

**Figure 1.** Overview of the analysis of this study. (a) Baseline (best model) selection. (b) Analysis 1 for evaluating the effect of expertise on the cross individual generalizability. (c) Analysis 2 for identifying spatial and temporal information that enhance cross-individual generalizability.

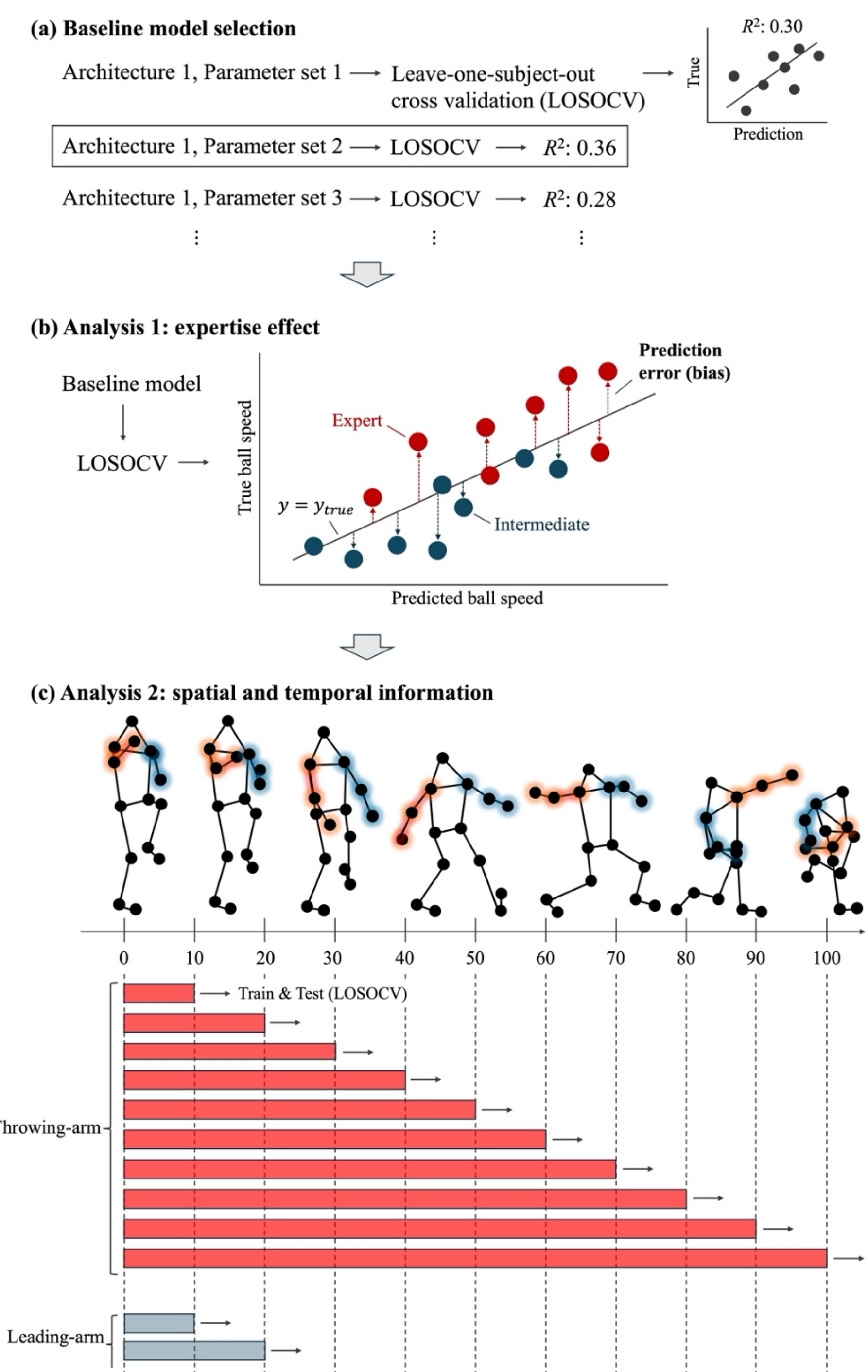

**Figure 2.** Scatter plot of the mean predicted and true ball speeds for individual pitchers. The results obtained with the selected baseline model are presented. Data points are color-coded according to a competitive level.

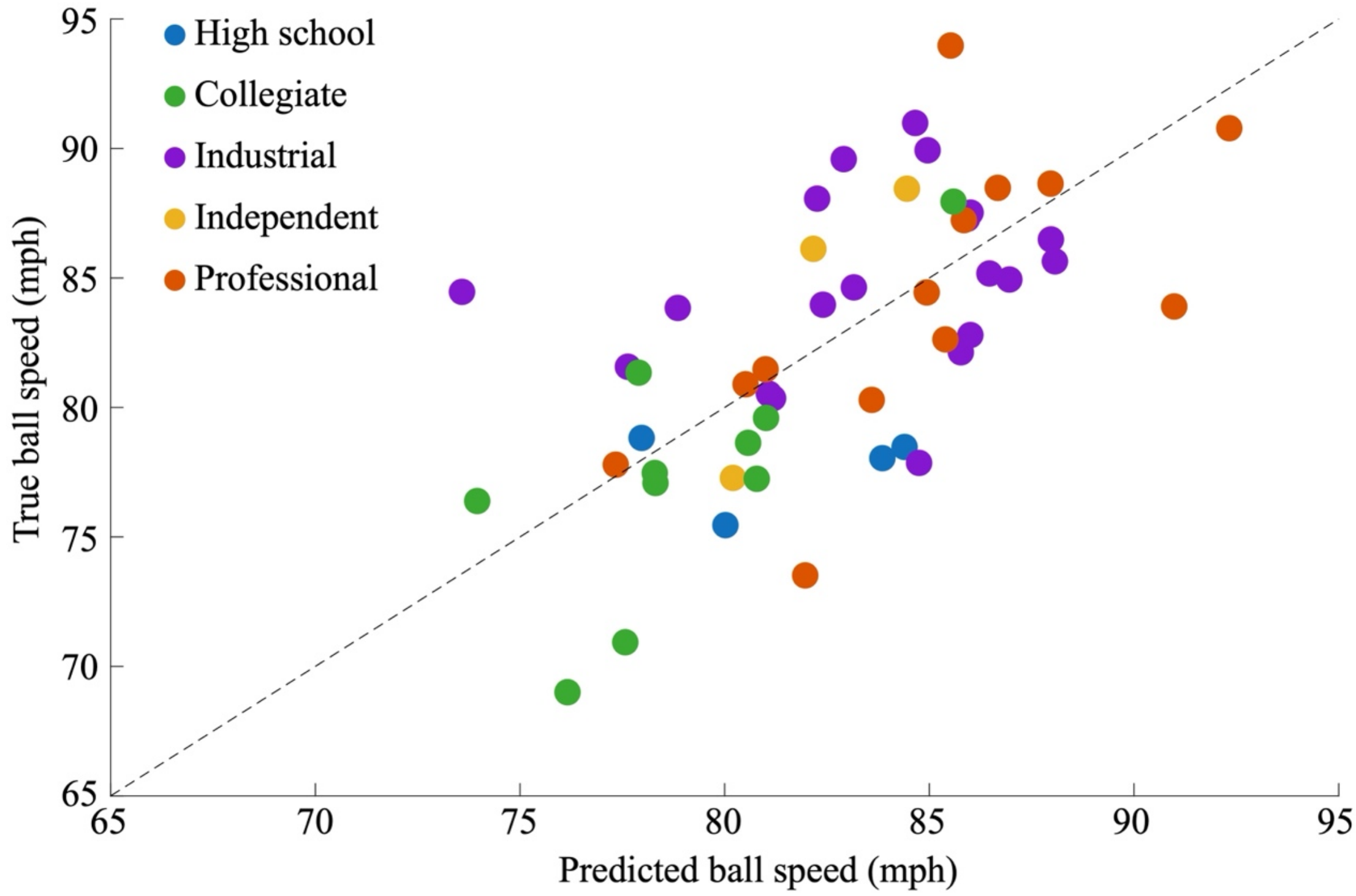

**Figure 3.** Results of Analysis 1. (a) Mean absolute prediction error and standard deviation for each group. (b) Mean signed prediction error and standard deviation for each group (*: p < .05).

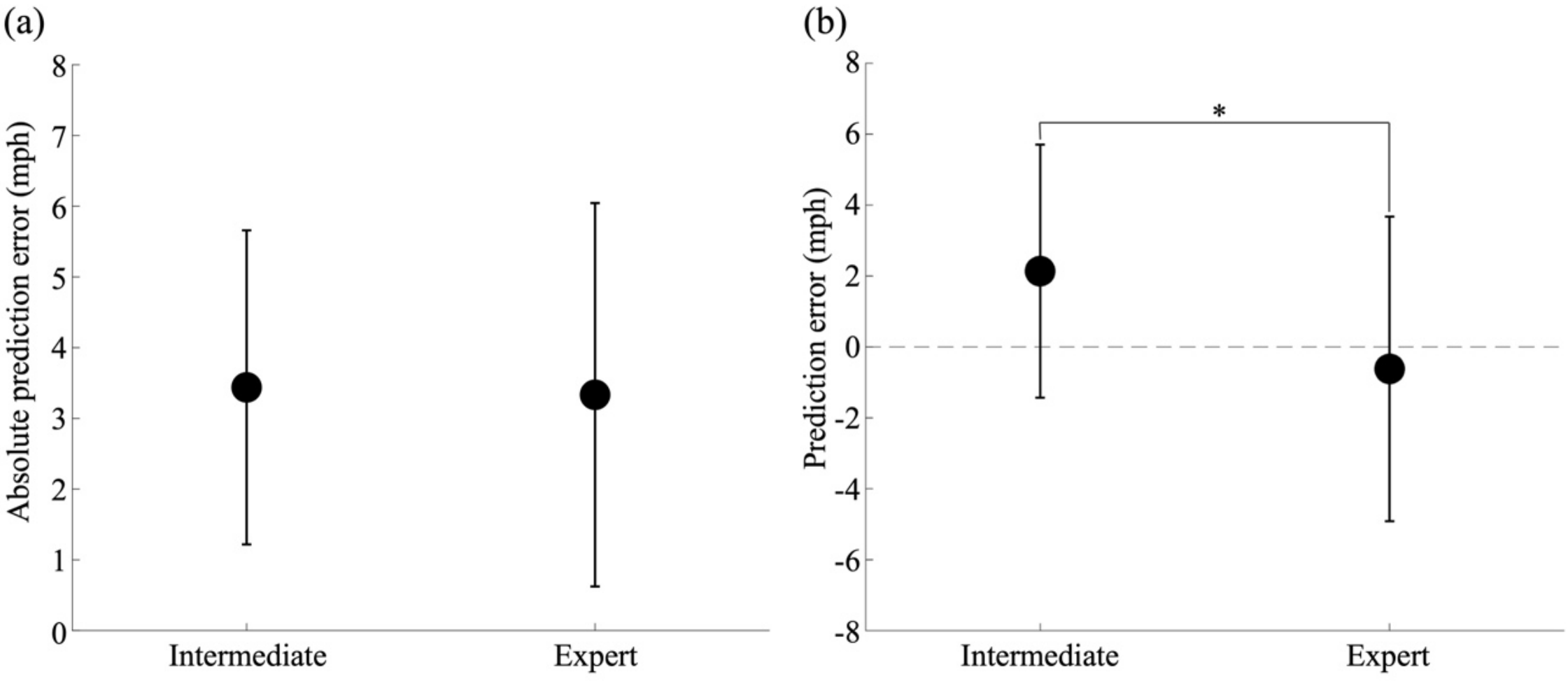

**Figure 4.** Results of Analysis 2. The mean and standard deviation of the generalization performance for each body region at each time span.

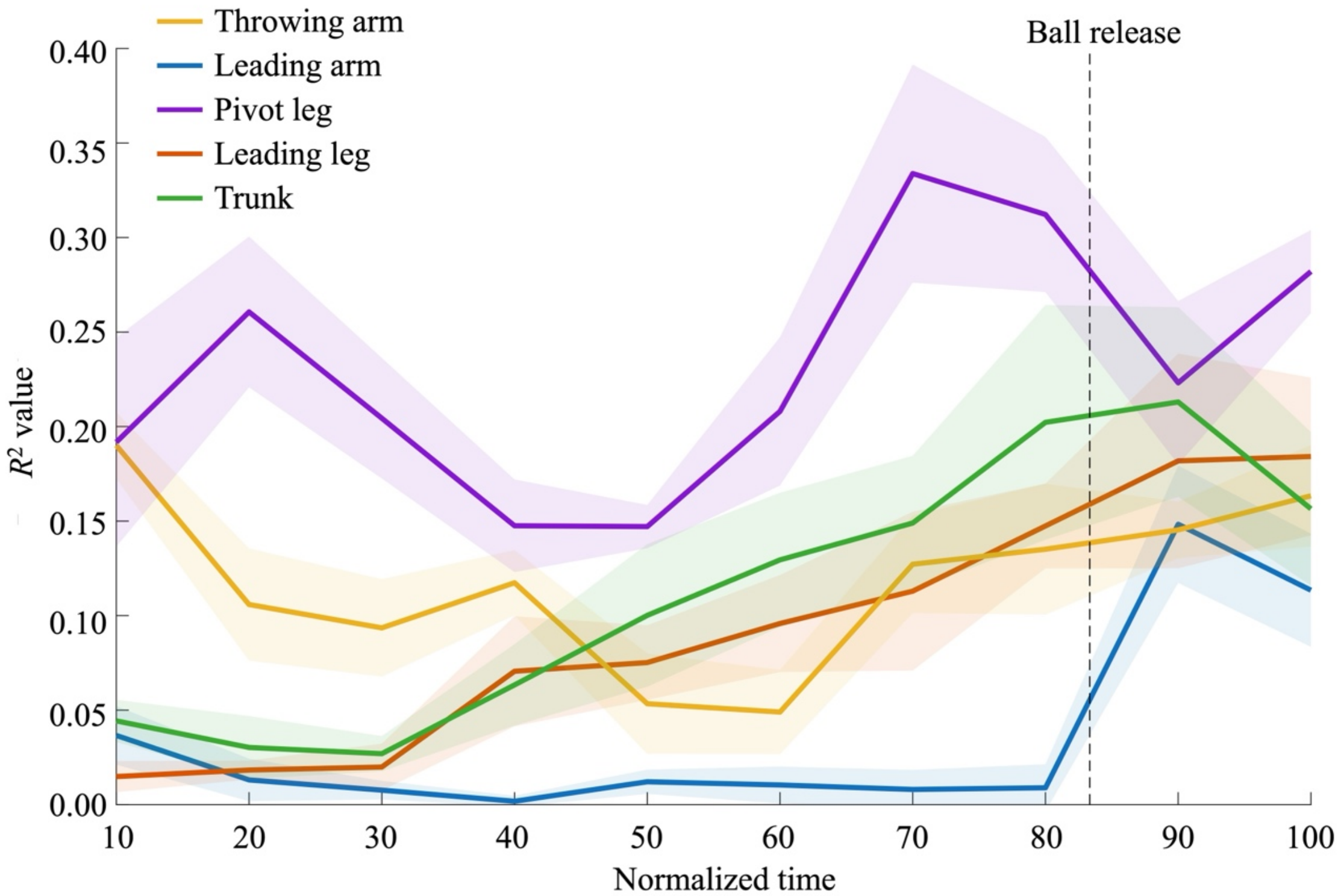

**Figure 5.** Example postures of (a) Intermediates and (b) Experts at the 20% time point. For each group, representative participants from the top 10% and bottom 10% in ball velocity are shown.

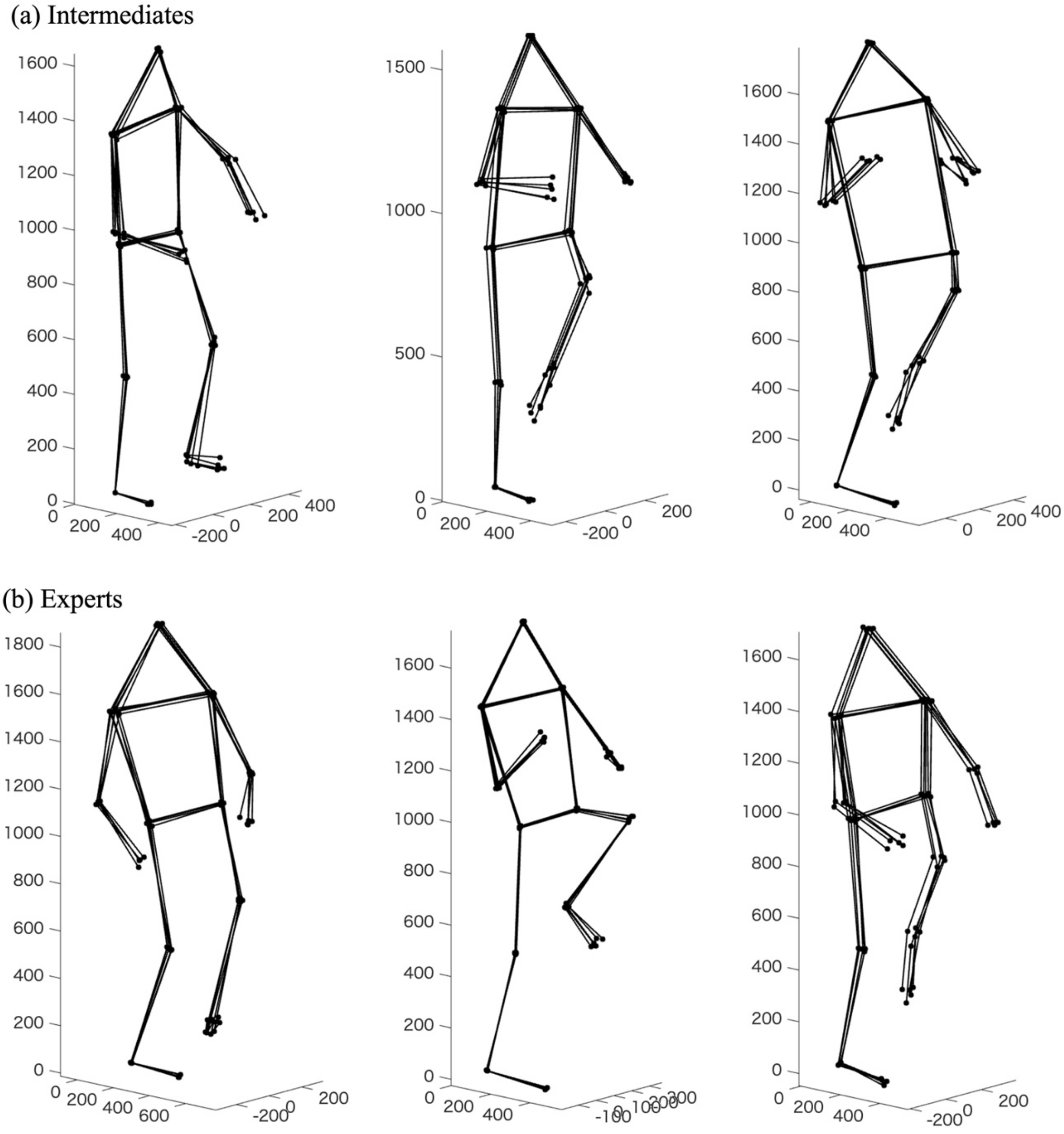